\documentclass{ws-p8-50x6-00}
\usepackage{epsf}
\begin{document}

\title{Color and Kinematic Interference Influence on the Probability of Color
Singlet Chain States}

\author{Yi Jin, Qubing Xie}

\address{Physics Department, Shandong University\\
Jinan Shandong 250100, People Republic of China\\ 
E-mail: tpc@sdu.edu.cn~~~~~xie@sdu.edu.cn}


\maketitle

\abstracts{
In popular event generators of high energy reaction, the 
large Nc, the number of color approximation is implied, which reduces 
the possible interference effects, and the probability of color singlet chain 
reaches to about $100\%$. In $Nc=3$ real world, we show that not only the color 
interference decreases this probability to $83\%$, $77\%$, $72\%$, $\cdots$, 
respectively 
to $n=2$, 3, 4, $\cdots$, \mbox{in the process of }$e^+e^-{\longrightarrow}q\bar{q}+ng$ 
final state, but also 
the accompanying kinematic interference exists, which decreases this 
probability to about $67\%$, $58\%$, $\cdots$, further. In the meanwhile, we find that the 
kinematic interference makes the probability of color separate singlet states 
increases synchronously. Combing and analyzing above facts, we infer that 
the probability of color singlet chain states 
in high energy ${e^+e^-}{\longrightarrow}{h's}$ reaction would be much smaller 
than $100\%$, which is commonly accepted. And the fraction of other color connection states, $e.g.$, color separate singlet states should be significant.}
\section{Introduction}
Color singlet chain states stand for a kind of way of color connection. In $q\bar{q}+ng$ system, it begins at quark, 
connects gluons one by one, and ends at antiquark. Their color states can be 
written as{\cite{ref1}
\begin{equation}\label{eq1}
\begin{array}{lll}
\lbrace{|f_i\rangle,i=1,2,3,{\cdots},n!}\rbrace & {\equiv} & \\ & & \lbrace{Nc^{-(n+1)/2}{|1_{qP(1)}1_{P(1)P(2)}{\cdots}1_{P(n)\bar{q}}}\rangle}\rbrace
\end{array}
\end{equation}

In popular generators, such as JETSET, a large Nc approximation is implied, so the probability of color singlet chain states come to one and they become the only way of color connection. But we know that in the real world, Nc is 3. It leads that the probability of color singlet chain states is no longer 1, and now it is necessary to calculate the probability in detail. 

To do this will make us encounter two types of interference, color interference and kinematic interference. These interferences will not come up when large Nc approximation. When Nc is limitted, any two color singlet chain 
states are not orthogonal to each other{\cite{ref1}} and this is the origin of color interference. Kinematic interference originates from
the variation of momentum structure under the condition of identical color 
connections. It has two very important nature:1. connected with color factors or color interference.  2. not the interference between Feynman diagrams, but contains the contribution of all Feynman diagrams. We will come to them later. Basicly, these two types of interference come from Nc-limit.
\section{Color Effective Hamiltonian Method}
To study the probability of color states for $q\bar{q}+ng$ system, a new method 
is proposed at tree level, which is called color Effective Hamiltonian 
method{\cite{ref2}}. 
By using color effective Hamiltonian $H_c$ constructed in [2] we can compute the amplitude of any color singlet 
\begin{equation}\label{eq2}
\langle f | H_c | 0 \rangle=\sum_P(\frac{1}{\sqrt{2}})^n{\langle{f|Tr(QG_1G_2{\cdots}G_n)^P}\rangle}D^P
\end{equation}
This expression can tell us the origion and nature of kinematic interference: First, from the color part of $\langle{f|Tr(QG_1G_2{\cdots}G_n)^P}\rangle$ 
we can get color factors. Then under the condition of identical color connection order of state $| f \rangle$, 
the momentum of gluons interchanging will cause different kinematic structure of $D^P$, and then different kinematic structure will cause kinematic interference $\sim Re(D^P{\cdot}D^{P'\ast})$ further when square of amplitude of Eq.~(\ref{eq2}).
\section{Color Interference}
We know that color singlet chain states are not orthogonal to each other when Nc limitted. So only after orthogonalization, 
can we calculate the cross section of color singlet 
chain states, and then the probability of them. The expressions of probability of SCS in two-gluon case is\cite{ref1}
\begin{equation}\label{eq3}
P_{SCS}=\frac{{\sigma}_{SCS}}{{\sigma}_{total}}=\frac{\int{d\Omega}{\lbrack4.45({|D^{12}|}^2+{|D^{21}|}^2)-3.1Re(D^{12}{\cdot}D^{21\ast})\rbrack}}
{\int{d\Omega}{\lbrack{\frac{\displaystyle 16}{\displaystyle 3}}({|D^{12}|}^2+{|D^{21}|}^2)-{\frac{\displaystyle 4}{\displaystyle 3}}
Re(D^{12}{\cdot}D^{21\ast})\rbrack}}
\end{equation}

The results for three gluon-case, four-gluon case and so on, have the similar form. When neglecting kinematic interference terms, after analytic calculation, we get
for $n=2$, $P_{SCS}\approx{4.45/(16/3)}\approx{83\%}$; $n=3$, $P_{SCS}\approx{5.47/(64/9)}\approx{77\%}$; for $n=4$, $P_{SCS}\approx{6.83/(256/27)}\approx{72\%}$. 
Here are some inferences according to the above:
$P_{SCS}\approx58\%~(\langle n\rangle=7(91GeV))$, $P_{SCS}\approx46\%~(\langle n\rangle=10(200GeV))$, $P_{SCS}\approx28\%~(\langle n\rangle=17(1TeV))$.
Where $\langle n\rangle$ is average gluon number for increasing energy, which is from JETSET74 with default parameters. These results indicate the decent trend of probability with {\sl n}. And we show that in Fig.~1.

Here we can give a naive physical understanding. We know that the probability for two gluons to have identical color is{\cite{ref3}} 1/8, then the probability for them to have different color is 7/8, {\sl i.e.}, the probability for them to form color singlet chain state is $\sim \frac{7}{8}$, for three gluons it is $\sim {(7/8)}^2$, similarly, for {\sl n} gluons, it is $\sim {(7/8)}^{n-1}$. Then we can say that the probability is reduced by $\sim (7/8)$. Qualitative analysis also shows that as long as Nc is 3, when {\sl n} increases, there will be more and more ways for selecting color connection among partons, especially color singlet with gluonic subsinglet(CS\cite{ref2}}), so the decence of probability of SCS is inevitable outcome. 
\begin{tabular}{ll}
\begin{minipage}{5cm}
\psfig{file=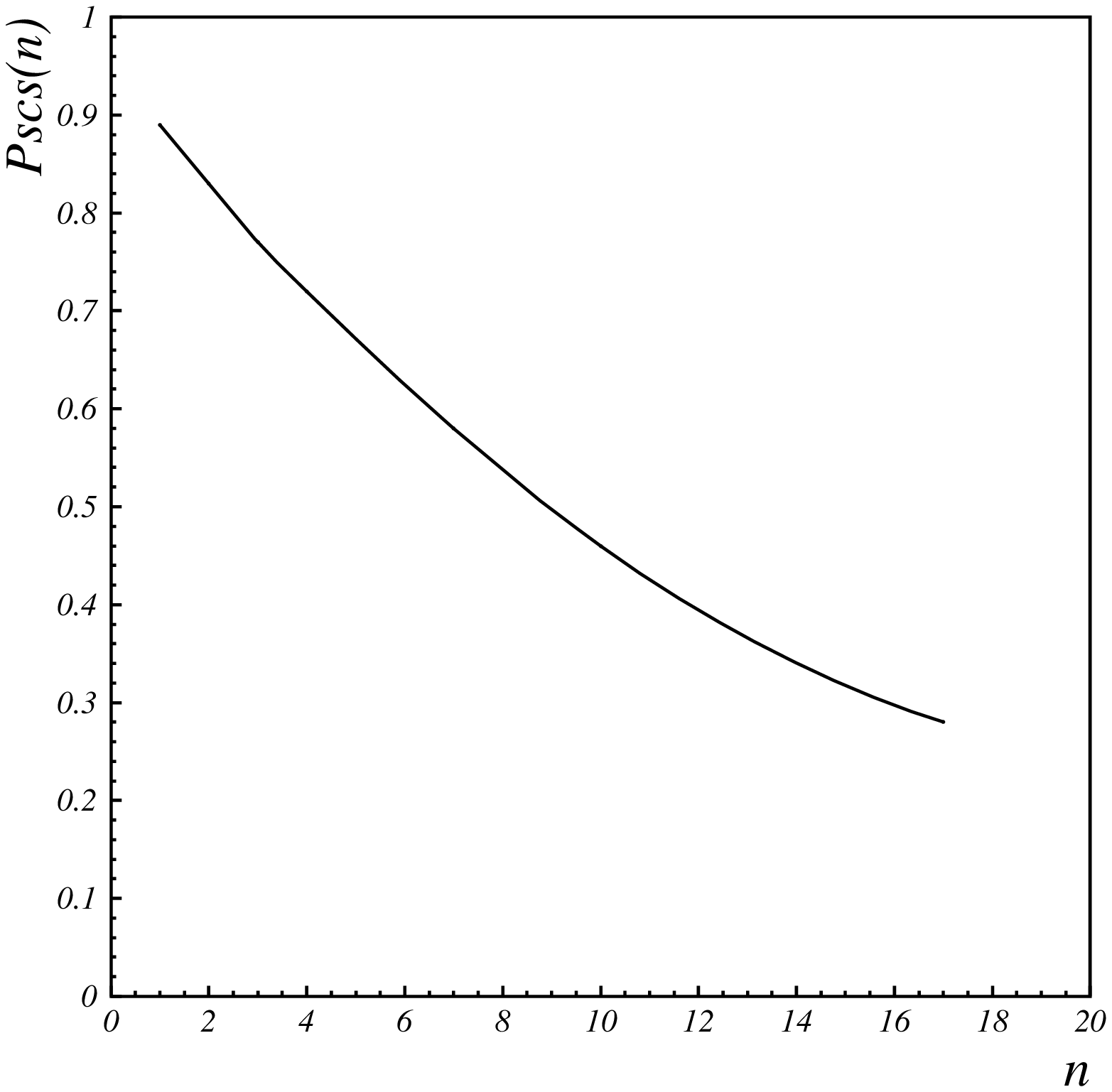,height=4cm}
\end{minipage}
&
\begin{minipage}{5cm}
Fig.1: Decent trend of $P_{SCS}$ with {\sl n} increasing
\end{minipage}\\[-0.5cm]
\end{tabular}
\section{Kinematic Interference}
Because kinematic interference contains the contribution from all Feynman diagrams, we need to know its expression to study it. We find there are two ways for getting the expression of kinematic interference terms. The first way is direct and relatively exact for concrete Feynman diagrams computing. But when {\sl n} is larger, the second way is convenient. By using soft-gluon-approximation, color effective hamiltonian is simplified to a form which contains the expression of $D^P$. We can find it in [1].

But we are not sure that if the results from two different ways have much difference. So for two-gluon case, we compared the results from them separately. The results is in Fig.~2. Where the solid line up in the fig is $P_{SCS}$ from concrete Feynman diagram computing; the dashed line is $P_{SCS}$ from soft-gluon-approximation. Fig.~2 shows that there is little difference between two ways. This implies that the second way is also valid and we can use it on three-gluon case. The result is in this Fig.~3.

We also study the probability of CS for two-gluon case. When two gluons have identical colors, they can form a subsinglet. The probability is{\cite{ref3}}
\begin{equation}\label{eq6}
P_{CS}=\frac{\int{d\Omega}{\lbrack{\frac{\displaystyle 2}{\displaystyle 3}}({|D^{12}|}^2+{|D^{21}|}^2)+{\frac{\displaystyle 4}{\displaystyle 3}}Re(D^{12}{\cdot}D^{21\ast})\rbrack}}
{\int{d\Omega}{\lbrack{\frac{\displaystyle 16}{\displaystyle 3}}({|D^{12}|}^2+{|D^{21}|}^2)-{\frac{\displaystyle 4}{\displaystyle 3}}Re(D^{12}{\cdot}D^{21\ast})\rbrack}}
\end{equation}
The results is the line below  in Fig.~2. 

There is still one thing we should note, here because color effective Hamiltonian is at tree level, our results must have the relation with cutoff. Throughout studying Feynman diagrams of two-gluon case, we find that when cutoff becomes small, the diagrams containing tri-gluon vertex make dominant contribution to total cross section of $e^+e^-{\longrightarrow}q\bar{q}+2g$. We know that these diagrams can not form CS because the two gluons of final parton state is coming from one mother-gluon propagator, then $q\bar{q}+2g$ can only form SCS. We can see in Fig.~2 and Fig.~3 that $P_{SCS}$ increases with cutoff decreasing.

\begin{tabular}{ll}
& \\[-0.6cm]
\begin{minipage}{5.6cm}
\psfig{file=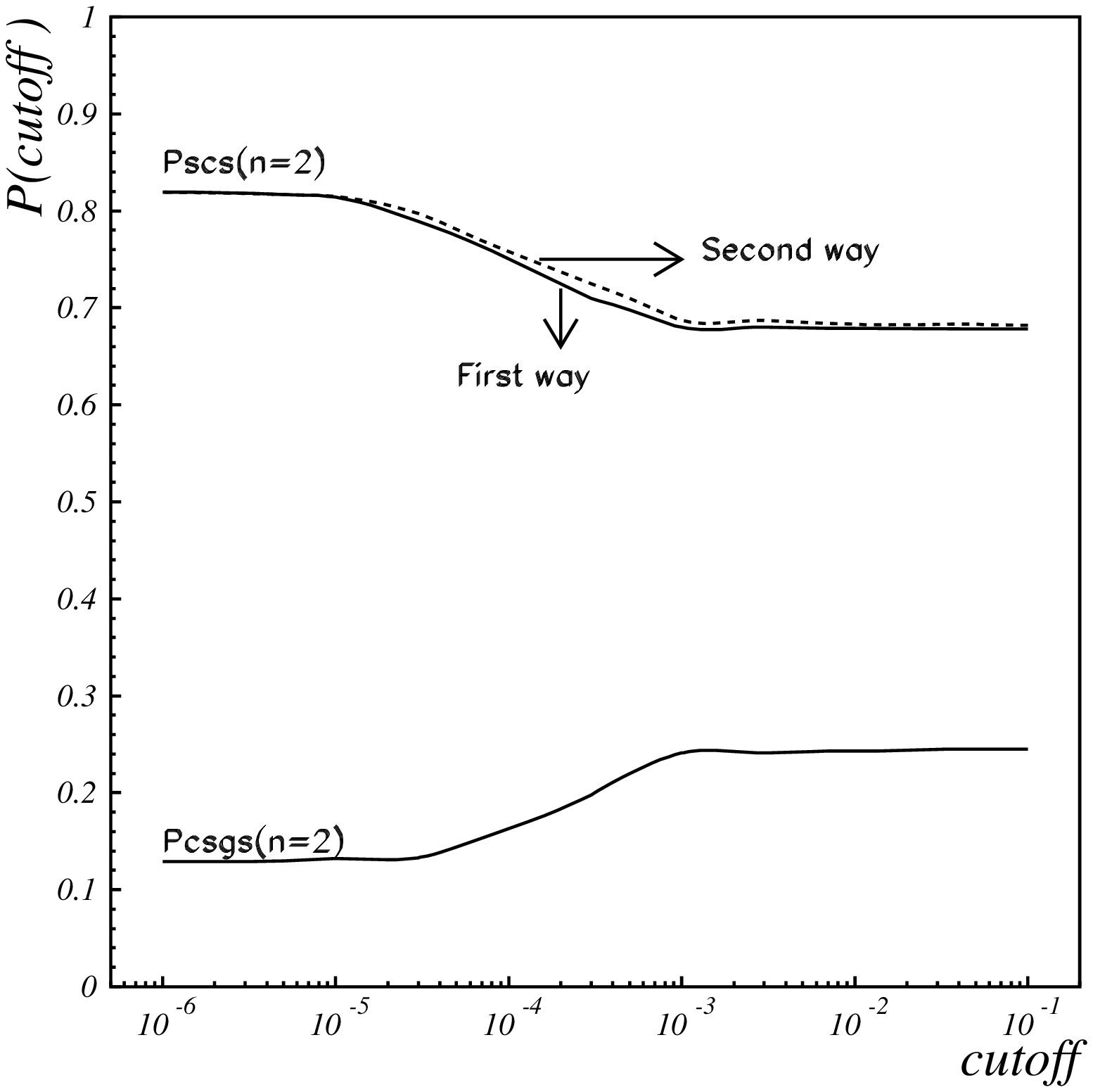,height=4cm}
\end{minipage}
&
\begin{minipage}{4cm}
\psfig{file=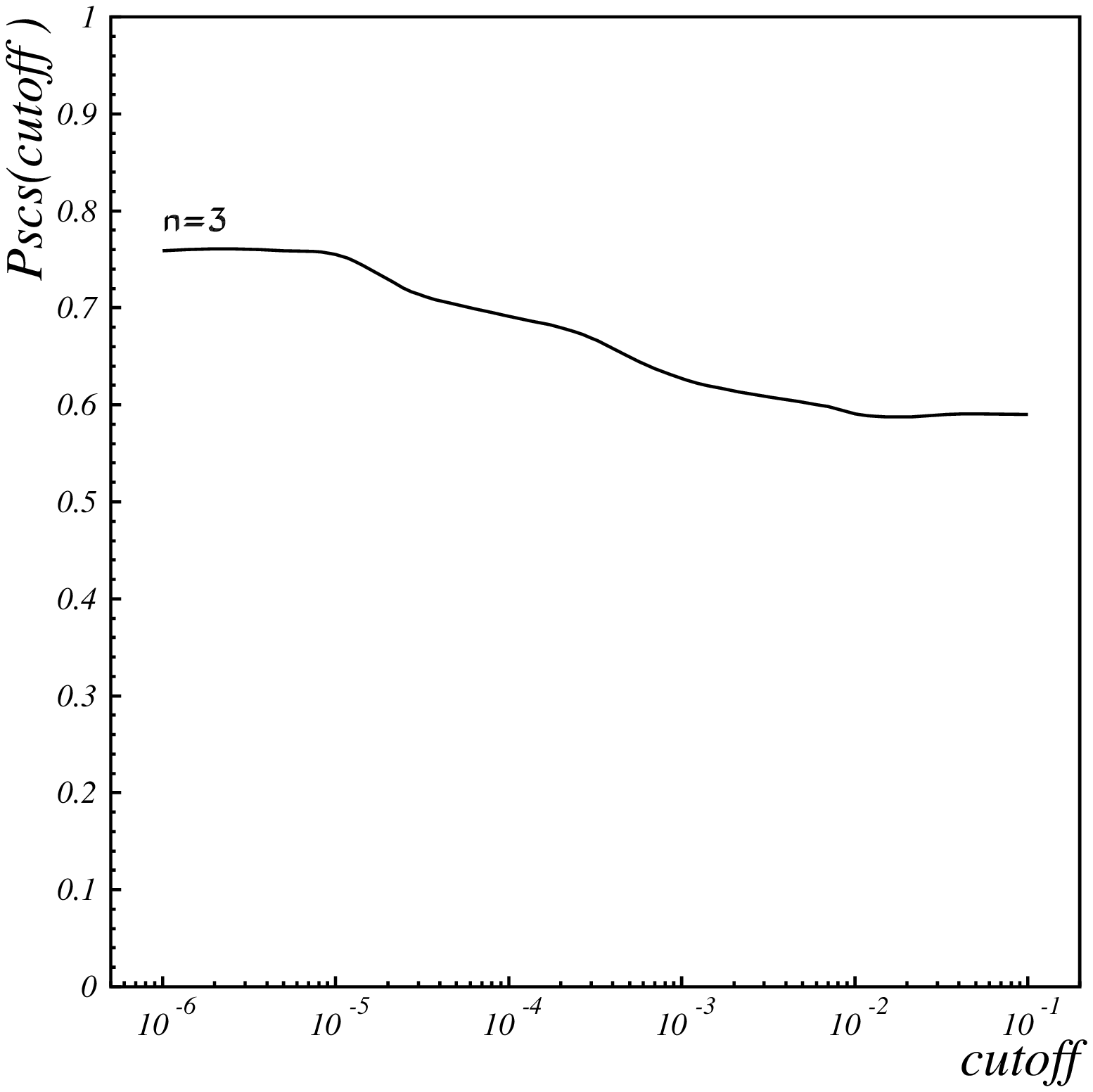,height=4cm}
\end{minipage}
\\[0.2cm]
\begin{minipage}{4.6cm}
Fig.2: 
Comparision of the results of $P_{SCS}$ for two-gluon case between two different ways. A line below the Fig is $P_{CS}$ for two-gluon case.
\end{minipage}
&
\begin{minipage}{5cm}
Fig.3: 
$P_{SCS}$ for three-gluon case including kinematic interference with \sl{cutoff}, using soft-gluon-approximation.
\end{minipage}\\[1.20cm]
\end{tabular}

These results indicate that kinematic interference makes the probability of color singlet chain states decent further; but make that of color singlet with gluonic subsinglet increasing.

When {\sl n} comes to 10, although the probability of SCS is less than 50\% and is no longer the main color connection ,can we still not say which color connection way, such as SCS and CS, is more reasonable and chosen by nature. Because the results above come from PQCD. Hadronization process is determined by perturbative phase and non-perturbative phase together. To find the truth, we still need to study hadronization results from different ways of color connection and the experiments. This will be discussed in [4].


\begin{thebibliography}{99}
\bibitem{ref1}Qun Wang, Gosta Gustafson, Yi Jin, Qu-bing Xie, Phys.Rev.D64, 012006(2001).
\bibitem{ref2}Qun Wang, Gosta Gustafson, Qu-bing Xie, Phys.Rev.D62, 054004(2000).
\bibitem{ref3}Qun Wang, Qu-Bing Xie, Zong-Guo Si, Phys.Lett.B388, 346(1996) .
\bibitem{ref4}Li Shi-yuan, Shao Feng-lan and Xie Qu-bing, Talk at this conference.
\end{thebibliography}
\end{document}